\documentclass[prl,twocolumn,showpacs,superscriptaddress]{revtex4-1}
\usepackage{graphicx,amssymb,amsmath,color,psfrag,bbm}
\begin{document}





\title{Nematic, topological and Berry phases when a flat and a parabolic band touch}
\author{Bal\'azs D\'ora}
\email{dora@eik.bme.hu}
\affiliation{Department of Physics and BME-MTA Exotic  Quantum  Phases Research Group, Budapest University of Technology and
  Economics, Budafoki \'ut 8, 1111 Budapest, Hungary}
\author{Igor F. Herbut}
\affiliation{ Department of Physics, Simon Fraser University, Burnaby, V5A 1S6, British Columbia, Canada}
\affiliation{ Max-Planck-Institut f\"ur Physik komplexer Systeme, N\"othnitzer Str. 38, 01187 Dresden, Germany}
\author{Roderich Moessner}
\affiliation{ Max-Planck-Institut f\"ur Physik komplexer Systeme, N\"othnitzer Str. 38, 01187 Dresden, Germany}

\date{\today}

\begin{abstract}
A (single flavor) quadratic band crossing in two dimensions is known to have a generic instability towards a quantum anomalous Hall
(QAH) ground state for infinitesimal repulsive interactions. Here we introduce a generalization 
of a quadratic band crossing which is protected only by rotational symmetry.
By focusing on the representative case of a parabolic and flat band touching, which also allows for a  straightforward lattice realization,
the interaction induced nematic phase is found generally
to compete successfully with the QAH insulator, and to become the dominant instability in certain parts of the phase diagram already at weak
coupling. The full phase diagram of the model, together with its topological properties, is mapped out using a perturbative renormalization group, strong coupling analysis, the
mean-field theory. Interestingly, the Berry flux varies continuously in the single flavour limit with  various control parameters.
\end{abstract}


\maketitle

\emph{Introduction.}
Topological states of matter possess the unique property that some of their response functions are universal, and independent of the sample-dependent
 microscopic parameters, such as scattering rate, interactions strength etc. The early members of this family were the celebrated quantum Hall states,
but with the advent of the topological insulator (TI), \cite{hasankane} numerous relatives have recently emerged. The topological protection of
these materials  mostly arises from their specific band structure, deriving from a strong spin-orbit interaction. Application-wise, TIs hold the promise
to revolutionize spintronics, and to contribute to conventional and quantum computing.

It is interesting to contemplate different physical mechanisms that could lead to non-trivial topological properties. Several strategies other than  band
structure engineering from the material science do exist. Time-periodic perturbations allow for modifying the Floquet band structure\cite{lindner,cayssol},
this way  influencing the topological properties of materials in situ without altering their composition. Applying strain to alter the band structure seems
also feasible for a variety of materials.\cite{guinea}

The common theme in these ideas is nevertheless the direct modification of the single-particle band structure. Electron-electron interactions, however,
 can also produce the desired effect. Simple mean-field decoupling of the interaction can mimic an effective spin-orbit coupling, for example,
thus inducing a transition from a topologically trivial to a non-trivial phase~\cite{raghu, vozmediano,igorPRB, roy}.
 How precisely this happens, and how competitive the topologically non-trivial phases in general are, is an open question
which often calls for more elaborate analysis \cite{tanja}.
The uncertainties notwithstanding,  clear-cut answers are available for two dimensional systems
with Fermi points instead of the usual Fermi surface. In Ref. \cite{kaisun,kunyang},
for example, it was argued that a single quadratic band crossing, protected by time reversal and rotational symmetry, is unstable with respect to topological insulating phases. 

Here
we formulate a more general quadratic band crossing Hamiltonian in two dimensions, protected only by  rotational symmetry. 
 A single copy of such a band crossing contains all three Pauli matrices and thus naturally breaks time reversal symmetry and possesses a non-trivial Berry phase.
We focus on the representative case of a flat and parabolic band touching, although our results apply more generally (see below). We show that the nematic order, previously suppressed at weak coupling\cite{kaisun,kunyang}, now also becomes 
possible within a certain range of the
parameters, even though the broken  time reversal symmetry might have suggested a quantum anomalous Hall (QAH) type phase.
 In particular, our (continuum) model
features a metastable phase purely nematically ordered, and without the QAH effect,
in contrast to the standard result\cite{kaisun}. We also propose a simple lattice realization of our generalized quadratic
band crossing Hamiltonian which could serve as an atomic physics platform  for the experimental study of the competition between the 
 different interaction-induced phases considered here.

\emph{Model.}
 Our generalization of the two-dimensional quadratic band crossing Hamiltonian \cite{kaisun} has a rotationally invariant 
spectrum, featuring however all {\it three}  Pauli matrices,
\begin{gather}
H_0=-\frac{p^2}{4m'}\mathbb{I}-\frac{p^2c}{4m}\sigma_3  -\frac{s}{4m}\left[\sigma_1(p_x^2-p_y^2)+\sigma_22p_xp_y\right],
\label{hamefflow}
\end{gather}
with the parameters $c=\cos(2\alpha)$, $s=\sin(2\alpha)$. Although its spectrum is independent of $\alpha$, 
the topological properties are not.
Eq. \eqref{hamefflow}  exhibits a quadratic band crossing with generally unequal effective masses for $|m|\leq |m'|$~\footnote{Note that $|m|>|m'|$ implies the touching of two inverted or normal parabola.}.
We focus on the representative case with $m=m'$, when one of the bands becomes flat, but all of our results apply to the more general 
situation.

\begin{figure}[h!]
\psfrag{t1}[t][][1][0]{$\epsilon_0>0$}
\psfrag{t2}[t][][1][0]{$\epsilon_0<0$}
\psfrag{dd}[t][][1][0]{$|\epsilon_0|$}
\includegraphics[width=6.5cm]{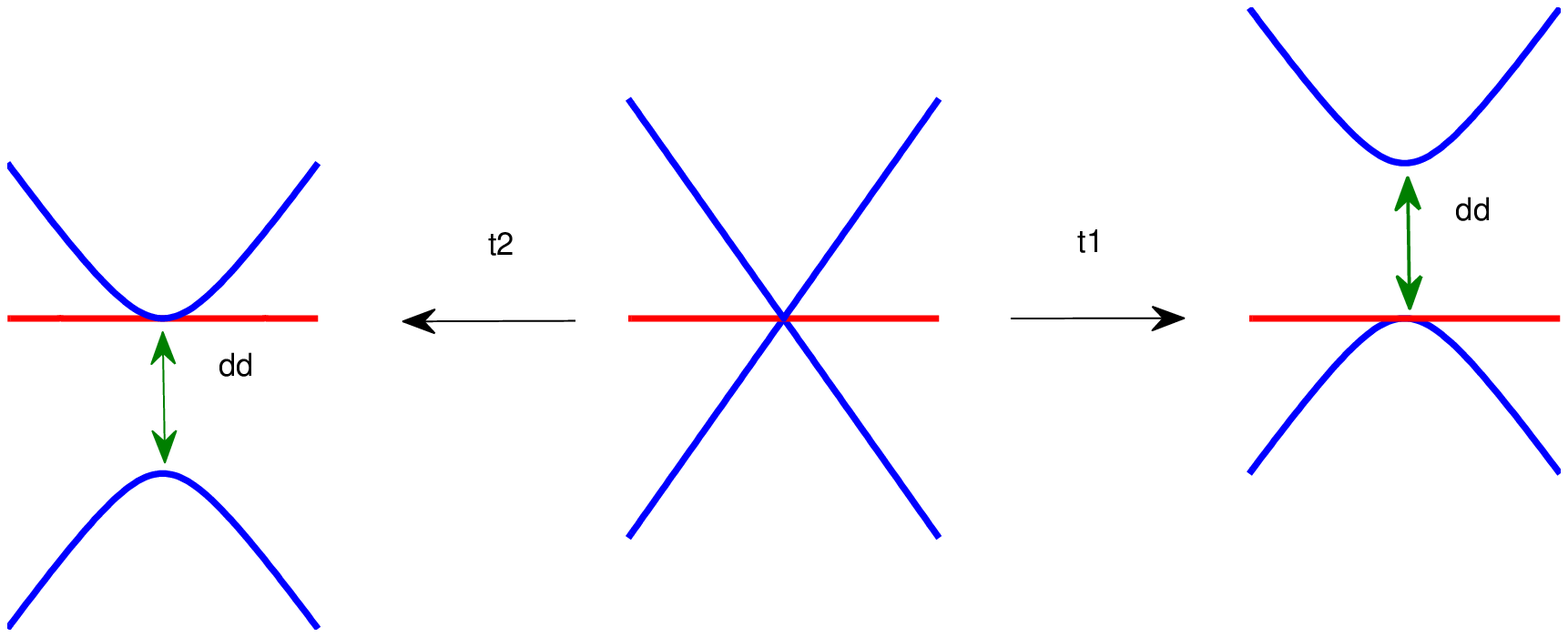}
\includegraphics[width=2cm]{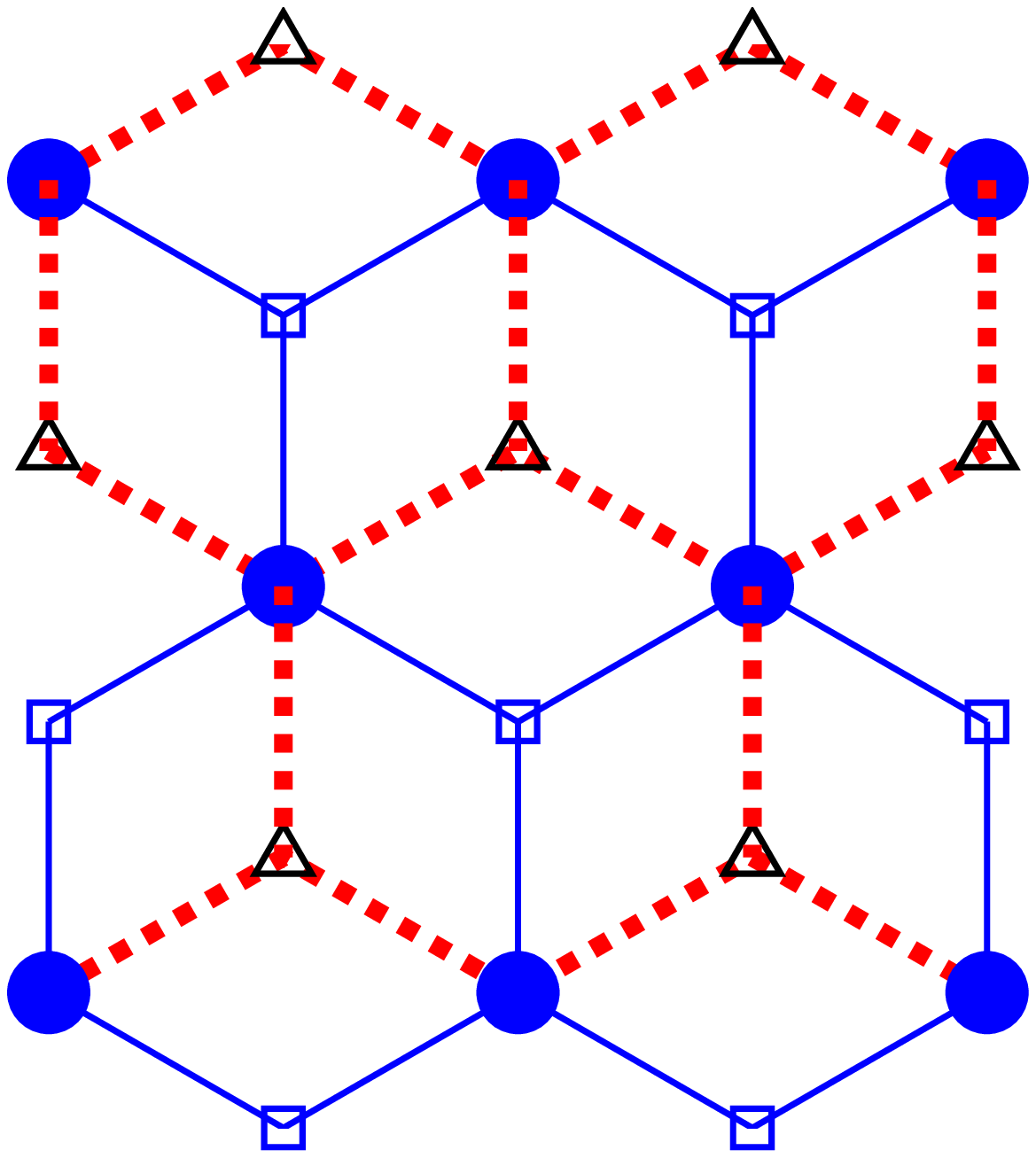}

\caption{(Color online) Left: The evolution of the non-interacting spectrum with $\epsilon_0$ is shown together with  the resulting quadratic and flat band touching. The horizontal red line
denotes the flat band, which remains fixed. Right: the dice lattice with $t\cos(\alpha)$ and $t\sin(\alpha)$ hopping along the blue solid and red dashed lines. The on-site energy of the six-fold connected filled blue sites, which are integrated out, is $\epsilon_0$.}
\label{specevol}
\end{figure}

 For $c=0$ ($\alpha=\pi/4$), this reduces to the model studied in Ref.~\cite{kaisun}.
The off-diagonal terms in Eq.~\eqref{hamefflow} are identical to those arising in the band-structure of the bilayer graphene. Due to the presence of all three Pauli matrices, the Hamiltonian necessarily violates the time-reversal symmetry. The spectrum consists of two bands,
one completely flat and another dispersive $\sim -p^2/2m$ (see Fig. \ref{specevol}).
For $m>0$, the low-energy dynamics is described by a filled inverted parabolic band touching an empty flat band at its maximum.
For $m<0$, a filled flat band touches an empty parabolic band at its minimum.
In spite of the explicit dependence of the Hamiltonian on the parameter $\alpha$ the spectrum of eigenvalues is independent from it. The eigenvectors of the flat and parabolic band, on the other hand, do depend on $\alpha$, as
$|0\rangle_p=\left[\sin(\alpha),-\cos(\alpha)\exp(2i\varphi_p)\right]^T$ and
$|p^2\rangle_p=\left[\cos(\alpha),\sin(\alpha)\exp(2i\varphi_p)\right]^T$.
In both cases, the flat band touches the parabolic band at a single point
in momentum space, producing a Berry phase of $2 \pi(1\pm c)$
for the flat and parabolic bands, respectively, that depends continuously on $\alpha$.
These non-$\pi$-quantized Berry phases indicate a no-go theorem for the Hamiltonian $H_0$, and require an even number of band touchings  as described by   Eq.~\eqref{hamefflow}, similarly to the Dirac equation in graphene. The exception is when $c=0$ or $s=0$, when
some of the three Pauli matrices are absent, as in Refs.~\cite{kaisun,vafek1}. In what follows we will focus on the case $m<0$, but our results can directly be
translated to the case $m>0$ as well.

Eq.~\eqref{hamefflow} remains invariant upon shifting $\alpha$ by $\pi$, and the ground state properties of the system are even functions of
$\alpha$, since its sign change can be compensated by a $\pi/2$ rotation of  the momentum. It suffices therefore to focus on $\alpha$ in the interval $[0,\pi/2]$.
The ``sublattice" symmetry is naturally broken in Eq. \eqref{hamefflow}, unless $\alpha=\pi/4$, since $\langle \Psi^+_1\Psi_1\rangle_0=\rho_0W\sin^2(\alpha)$ and
$\langle \Psi^+_2\Psi_2\rangle_0=\rho_0W\cos^2(\alpha)$ in the non-interacting ground state. Here, $\Psi_{1,2}$ are the field operators for the two species of electrons,
$W$ is the high energy cutoff, and $\rho_0=|m|/2\pi$ is the constant density of states in the parabolic band. The system is thus naturally a ``charge density wave" when $\alpha\neq \pi/4$.

Quadratic Hamiltonians resembling ours, but lacking the third Pauli matrix, arise in an effective collinear spin-density-wave theory\cite{chern},
from the surface states of certain Weyl semimetals,\cite{xu2011} as well as from the Lieb lattice.\cite{tsai}

Next, we define the full (interacting) low-energy Hamiltonian:
\begin{gather}
H=\int d{\bf r} \left[{\bf \Psi}^+({\bf r})H{\bf \Psi}({\bf r})+U\delta n_1({\bf r})\delta n_2({\bf r})\right],
\label{hint}
\end{gather}
where the spinor ${\bf \Psi}^+=(\Psi_1^+,\Psi_2^+)$, $U$ is the strength of the coupling constant,and $\delta n_{l}=\Psi_l^+\Psi_l-\langle\Psi_l^+\Psi_l\rangle_0$, with $l=1$, 2, stands for the densities measured from their non-interacting values.
The second term can be regarded as a fine-tuning of the interaction, which can be provided by single-particle terms of the form e.g. $\Psi_1^+\Psi_1 \langle\Psi_2^+\Psi_2\rangle_0$. Without this careful subtraction of the non-interacting densities a constant self-energy $\propto U\sigma_3$ would be generated, gapping out the spectrum already at the level of Hartree approximation.  This effect is known to occur for the semi-Dirac points\cite{doramerging}, and here comes as a consequence of the broken sublattice symmetry at a general $\alpha\neq \pi/4$ mentioned earlier.

\emph{Lattice realization.}
Eq.~\eqref{hamefflow} can describe a modified dice or $T_3$ lattice, consisting of three layers of triangular lattices with only intersublattice hoppings between adjacent layers, 
or equivalently,  two honeycomb lattices sharing one sublattice, shown in Fig. \ref{specevol}. The  Hamiltonian matrix reads\cite{bercioux,urban,diracdora}
\begin{gather}
H_{dice}=\left(\begin{array}{ccc}
0 & t_1f({\bf k}) & 0\\
t_1f^*({\bf k}) & \epsilon_0 & t_2f({\bf k})\\
0& t_2f^*({\bf k}) & 0
\end{array}\right),
\label{ham}
\end{gather}
where $t_1$ and $t_2$ are the hopping integrals between adjacent triangular lattices,  $\epsilon_0$ is an on-site potential in the middle layer (arising from e.g. a real chemical potential,  or from the Hartree term of a short range interaction) and $f({\bf k})=1+2\exp(i3k_y/2)\cos(\sqrt{3}k_x/2)$.
If $\epsilon_0$ is large compared to the energies of interest, the electrons on the middle layer can be integrated out \cite{vafek1} yielding the effective two-band Hamiltonian
\begin{gather}
H^{eff}_{dice}=-\frac{1}{\epsilon_0} \left(\begin{array}{cc}
|t_1f({\bf k})|^2 & t_1t_2f({\bf k})^2\\
t_1t_2f^*({\bf k})^2 & |t_2f({\bf k})|^2
\end{array}\right).
\label{hameff}
\end{gather}
One set of the eigenvalues of the effective Hamiltonian is a completely dispersionless flat band, whereas the other one reads as
$-(t_1^2+t_2^2)|f({\bf k})|^2/\epsilon_0$. Around the $K$ point in the Brillouin zone one can linearize the function $f({\bf K+p})\approx (3/2)(p_x-ip_y)$, and upon further parameterizing the hopping integrals  in terms of an ``angle" $\alpha$ as $t_1=t\cos(\alpha)$, $t_2=t\sin(\alpha)$, the low energy dynamics is described by Eq.~\eqref{hamefflow} with $m=2\epsilon_0/9t^2$, $t=\sqrt{t_1^2+t_2^2}$. Similarly as in graphene, a time-reversed copy of the Hamiltonian with $\sigma_2\rightarrow -\sigma_2$ describes the low energy physics at the opposite corner of the hexagonal Brillouin zone, at the point $K'$.
The two Hamiltonians at the $K$ and $K'$ points map onto each other under time reversal, and therefore taken together preserve the time-reversal symmetry.
Nevertheless, for the sake of simplicity, in studying the effects of interactions our focus will mainly be on a single valley.

\emph{Weak coupling analysis.}
The model at hand shares similarities with those for bilayer graphene\cite{kunyang,zhangblg,lemonik} and the simpler quadratic band touching in two dimensions\cite{kaisun}. The dynamic critical exponent is $z=2$, which together with the spatial dimensionality of 2 predicts short range interactions to be precisely marginal at the tree level. By performing the standard one-loop renormalization group (RG), assuming the flat band to be fully filled and the parabolic band to be empty, the repulsive interactions turn out to be marginally {\it relevant}. In fact, exactly as in the previously studied case of a single-valley quadratic band crossing,\cite{kaisun} the particle-particle diagram vanishes and only the particle-hole diagrams contribute. This can be understood as a consequence of the absence of the ``fermionic" (negative when squared) time-reversal symmetry necessary for Cooper pairing\cite{igorprd} in both cases. The resulting beta-function is then
\begin{gather}
\frac{dU}{d\ln s}=U^2\rho_0 + O(U^3),
\label{eq:rg}
\end{gather}
which is, unexpectedly, completely independent of the parameter $\alpha$, at least to the leading order.
Here we integrated out the fermions with momenta within the shell
$[ W/s, W ]$ and with all Matsubara frequencies. To the leading order in $U$ neither the angle $\alpha$ nor the mass in Eq. (1) flow.

\begin{figure*}[t!]
\psfrag{x}[t][][1][0]{$4\alpha/\pi$}
\psfrag{y}[b][][1][0]{$\Delta/|\mathcal M|$}
\includegraphics[width=4.3cm]{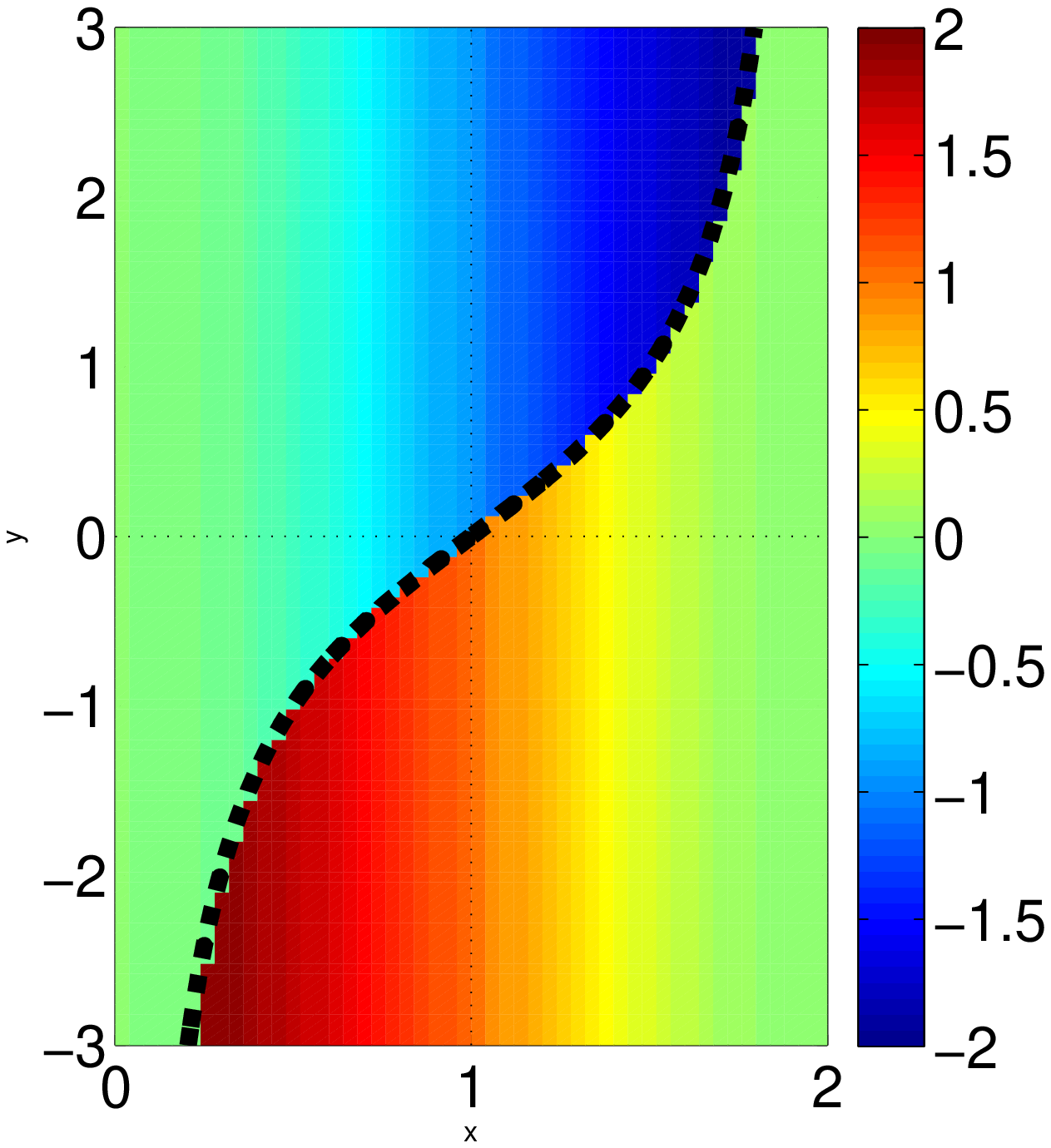}
\hspace*{2mm}
\psfrag{x}[t][][1][0]{$4\alpha/\pi$}
\psfrag{y}[b][][1][0]{$\Delta^K/|\mathcal M|$}
\psfrag{g1}[][][1.7][0]{\bf -2}
\psfrag{g2}[][][1.7][0]{\bf 2}
\psfrag{g3}[][][1.7][0]{\bf 0}
\includegraphics[width=3.5cm]{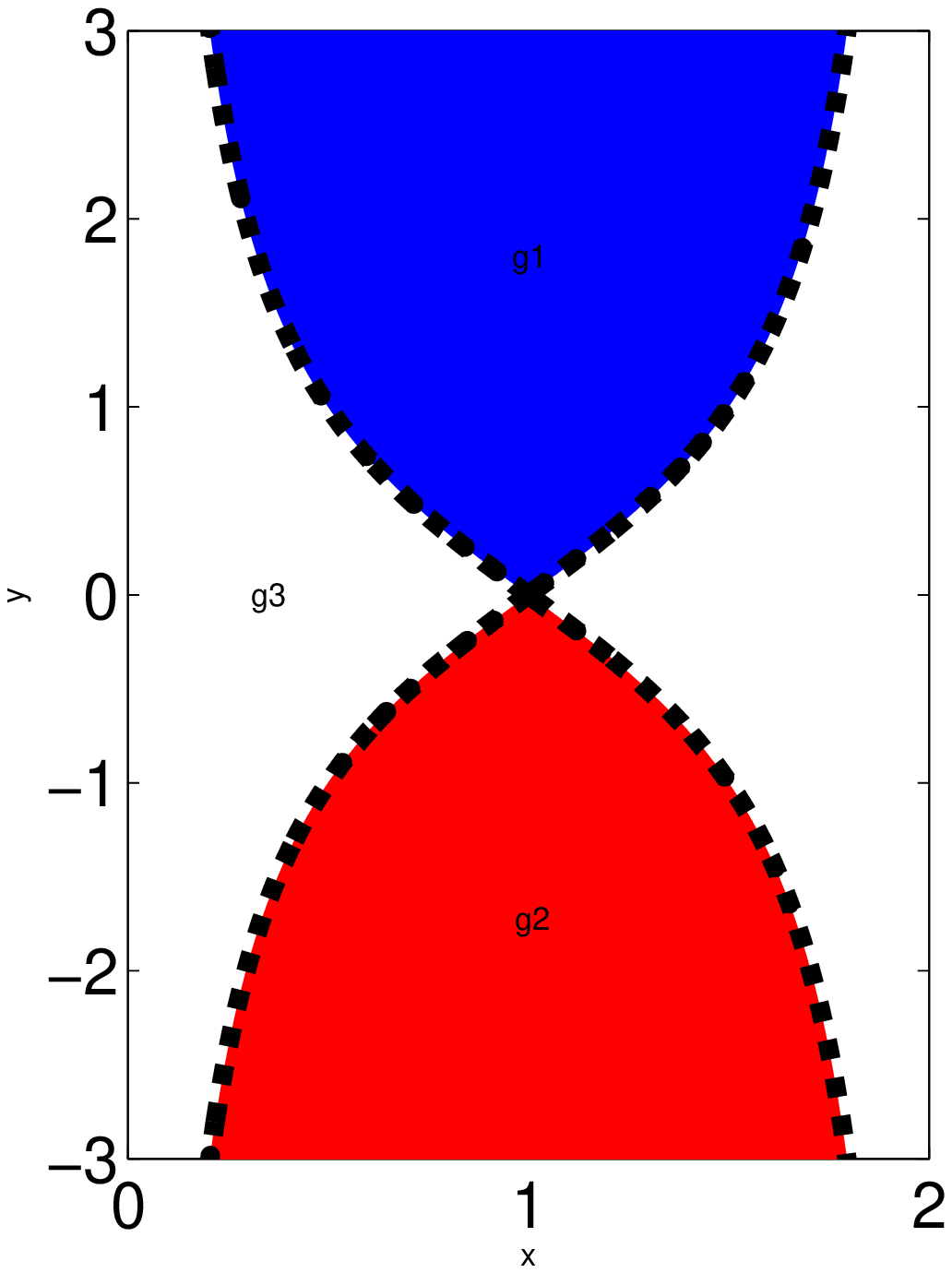}
\hspace*{2mm}
\psfrag{x}[t][][1][0]{$4\alpha/\pi$}
\psfrag{y}[b][][1][0]{$\Delta/W$ (blue), $\mathcal M/W$ (red)}
\psfrag{metastable}[][][1][0]{\textit{metastable}}
\psfrag{stable}[][][1.3][0]{\textbf{stable}}
\includegraphics[width=6cm]{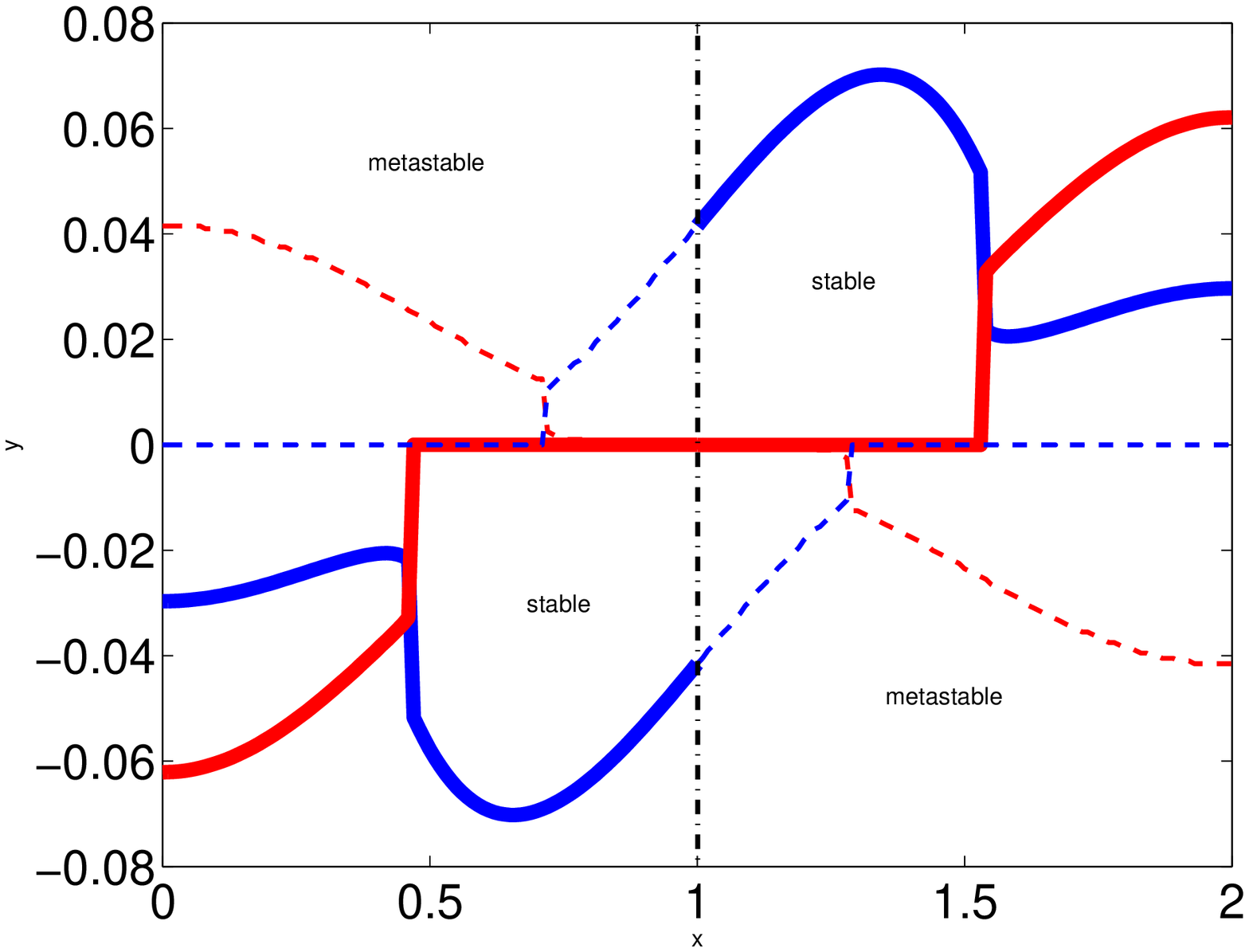}

\caption{(Color online) Left panel: the Berry flux for the single valley Hamiltonian in Eq.~\eqref{hamefflow} with QAH and nematic order parameters, becoming quantized only when $c$ or $s$ vanishes. The Berry flux jumps
when $\Delta/|\mathcal M|=-c/|s|$.
Middle panel: the Chern-number of the dice lattice with two time reversed
copies of  Eq.~\eqref{hamefflow} at the $K$ and $K'$ points, assuming $\Delta^K=-\Delta^{K'}$. The phase boundaries between phases with different Chern numbers are given by
$\Delta^K/|\mathcal M|=\pm c/s$.
For $\Delta^K=\Delta^{K'}$, the Chern number is identically zero.
The phase of the nematic order parameter has no effect on the  two left panels.
Right panel: the typical evolution of the order parameters is shown for $U\rho_0=0.31$. The sign of the QAH order is well defined (i.e. its sign change would alter the ground state energy)
since it is reached through a first order
phase transition for $\alpha\neq\pi/4$. At $\alpha=\pi/4$, the transition is second order and is identical to Ref.~\cite{kaisun}.
Only the absolute value of the nematic order parameter is fixed from the mean-field equations, its phase, $\theta$ is arbitrary anywhere on the phase
diagram.}
\label{phasediag}
\end{figure*}

In order to determine the type of ordering that ensues, we rewrite the interaction in a more suggestive form as
\begin{gather}
8 \Psi_1^+\Psi_1\Psi_2^+\Psi_2=\left({\bf \Psi}^+{\bf \Psi}\right)^2-\sum_{l=1}^3\left({\bf \Psi}^+\sigma_l{\bf \Psi}\right)^2,
\end{gather}
from which the orderings preferred by the repulsive interaction may be readily identified as
\begin{subequations}
\begin{gather}
\Delta=-\frac{U}{2}\left(\langle {\bf \Psi}^+\sigma_{3}{\bf \Psi} \rangle-\langle {\bf \Psi}^+\sigma_{3}{\bf \Psi} \rangle_0\right),\\
\mathcal M\exp(i\theta)=-\frac{U}{2}\langle {\bf \Psi}^+\left(\sigma_1+i\sigma_2\right){\bf \Psi}\rangle,
\end{gather}
\label{orderparameters}
\end{subequations}
where $\theta$ keeps track of the relative phase of the two nematic order parameters.
All of these yield a fully gapped spectrum for $\alpha\neq \pi/4$, and since Eq.~\eqref{hamefflow} contains all three Pauli matrices, they also feature a finite zero-field Hall conductivity, which is, however, not quantized in general.
This parallels closely the non-integer quantized Hall response of a single Dirac cone\cite{ludwig94}, which only gets quantized upon considering its time-reversal partner, similarly to graphene.
 The $\Delta$ corresponds to the QAH state, which gaps out the spectrum and does not break any additional symmetry, such as time reversal, when $\alpha\neq \pi/4$, since
all three Pauli matrices are already present in the bare Hamiltonian in Eq.~\eqref{hamefflow}. Only when $\alpha=\pi/4$ and the matrix
$\sigma_3$ is absent does the QAH phase break the time-reversal symmetry. On the other hand, $\mathcal M$ describes the nematic orderings,
which would results from the spontaneous reduction of the $C_6$ rotational symmetry down to $C_2$, with the full rotational symmetry of the spectrum
 broken (see Eq.~\eqref{mfspectrum}). When $c=0$,
the nematic phase becomes gapless with two linearly dispersing Dirac cones, similarly to Refs.~\cite{kunyang,kaisun}, and its zero-field Hall conductivity vanishes.
The Berry flux\cite{hasankane,diracdora}, characterizing the QAH and nematic orderings and calculated from the single valley realization, is shown in Fig.~\ref{phasediag}.

The ratio of the susceptibilities corresponding to these order parameters is $\alpha$-dependent:
 \begin{gather}
\frac{\chi_\Delta}{\chi_{\mathcal M}}=\frac{2s^2}{2-s^2}.
\end{gather}
Although the spectrum, and even the RG flow  of the interaction, were oblivious to the parameter $\alpha$, it nevertheless determines the
leading susceptibility, and thus the ultimate nature of the instability at weak coupling. The nematic and QAH susceptibilities are equal
only at a critical value $\alpha_c$, given by $\sin^2(2\alpha_c)=2/3$. The QAH state is realized when $|\sin(2\alpha)|>\sqrt{2/3}$ and
thus  $\chi_\Delta>\chi_{\mathcal M}$, with the nematic order being otherwise dominant.

In the case of two valleys, relevant to the dice lattice, the order parameter $\Delta$  with different signs in the two valleys would break time reversal
invariance and result in an overall finite QAH effect. In the case of identical signs, $\Delta$ would only additionally contribute to the amplitude of the
charge density wave, which already exists for general $\alpha\neq \pi/4$.
Albeit the nematic order parameter possesses a non-zero Berry flux in a single valley, the contribution from the other valley with opposite chirality always compensates it to zero, at least in the physically motivated case when the
absolute value of the the two order parameters in the two valleys are equal.
When QAH and nematic order coexist, the Chern-numbers are shown in Fig.~\ref{phasediag}.

\emph{Strong coupling analysis.}
 Without the kinetic energy term ($t=0$), the interaction clearly favors deviations from the particle densities in the non-interacting limit,
with either $\langle \delta n_1\rangle>0$ and $\langle \delta n_2\rangle<0$,  or vice versa. Depending on the particle densities in the non-interacting limit,
one of these  would be energetically favorable, and the ground state energy profile as a function of the respective particle densities develops an asymmetric double
well structure, leading to a first order transition. At $\alpha=\pi/4$, the depths of the two wells become equal, and the order of the transition changes from first
to second. As a result, the system will be a ``fully polarized" charge density wave in the sense that one sublattice is fully occupied while the other is empty.
This corresponds to $\Delta\neq 0$, i.e. the analog of the QAH state for all values of $\alpha$.

Connecting the weak and strong coupling regimes when $|\sin(2\alpha)|<\sqrt{2/3}$
requires therefore a quantum phase transition from the nematic to the QAH state with
increasing interaction. The details of this transition are evidently beyond the reach
of the weak-coupling RG calculations, and so we formulate a mean-field theory in order to study it further. For $|\sin(2\alpha)|>\sqrt{2/3}$, on the other hand, the same QAH state appears at both strong and weak couplings.

\begin{figure}[h!]
\psfrag{x}[t][][1.3][0]{$\alpha$}
\psfrag{one}[t][][0.9][0]{$\pi/4$}
\psfrag{0}[t][][0.9][0]{$0$}
\psfrag{pi}[t][][0.9][0]{$\pi/2$}
\psfrag{ac1}[t][][0.9][0]{$\alpha_c$}
\psfrag{ac2}[t][][0.9][0]{$\pi/2-\alpha_c$}
\psfrag{y}[][][1.3][0]{$U\rho_0$}
\psfrag{t1}[t][][1.4][-70]{\color{blue}\textbf{nematic}+\tiny QAH}
\psfrag{t2}[t][][1.4][70]{\color{blue}\textbf{nematic}+\tiny QAH}
\psfrag{t3}[t][][1][90]{\color{red}\textit{QAH}}
\psfrag{t4}[t][][2][0]{\color{blue}\textbf{QAH}}
\psfrag{t5}[t][][1][0]{\color{red}\textit{nematic}}
\psfrag{t6}[t][][1][0]{\color{red}\textit{nematic}}
\includegraphics[width=6cm]{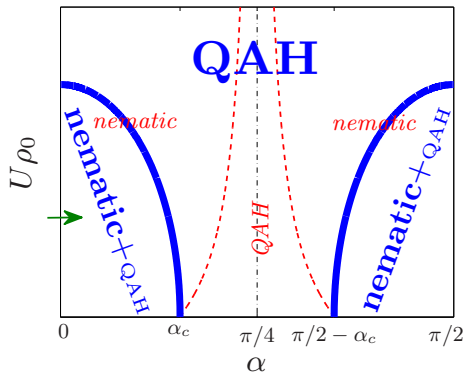}

\caption{(Color online) Schematic phase diagram in the $U-\alpha$ plane. The stable solutions are depicted in blue, while the metastable part is in red.
The transition is first order across the lines between the various phases on either side of the black dash-dotted line, along which the transition is
second order to a QAH state. Fig.~\ref{phasediag} shows the order parameters along a cut indicated by the green arrow.}
\label{phasediagall}
\end{figure}

\emph{Mean-field theory.}
Allowing for all three kinds of orderings, the mean-field decoupling of the interaction gives the energy spectrum
\begin{gather}
E_{\pm}({\bf p})=\frac{\varepsilon_p}{2} \pm\nonumber\\
\pm\sqrt{\frac{\varepsilon_p^2}{4}+\varepsilon_p(\mathcal M s\cos(2\varphi_p -\theta)+\Delta c)+\Delta^2+\mathcal M^2},
\label{mfspectrum}
\end{gather}
where $\varepsilon_p=p^2/2|m|$.

The ground state energy per unit cell is
\begin{gather}
E=\frac{\Delta^2+\mathcal M^2}{U}+\int \frac{d^2p}{(2\pi)^2}E_-({\bf p})+W\rho_0\Delta c,
\end{gather}
subject to minimization with respect to $\Delta$ and $\mathcal M$. The last term arises from the fine tuning of the interaction in terms of the non-interacting densities. The relative angle of the nematic order parameters
$\theta$ drops out from
the calculation. Were  the rotational symmetry of the spectrum already broken in Eq.~\eqref{hamefflow} by, for example, choosing unequal prefactors of the $\sigma_1$ or $\sigma_2$ term, the preferred value of $\theta$ would
also be determined by the mean-field equations, and the competition between different nematic orders and the QAH phase would be more subtle.

The Ginzburg-Landau expansion of $E$ contains even and odd powers of $\Delta$, therefore the energy landscape exhibits
an asymmetric double well structure, leading to a first-order phase transition. On the other hand, only even powers of $\mathcal M$ are present
in $E$, yielding a second order phase transition for the nematic order. Close to $\alpha\lesssim \pi/4$, a pure QAH state is stable for $U\rho_0\ll 1$, and
\begin{gather}
\Delta=\frac{W}{c-1}\exp\left(-\frac{1}{U\rho_0s^2}+\frac{c}{1-c}\right),
\label{deltawc}
\end{gather}
displaying the characteristic essential singularity in the weak coupling limit. For $\alpha\gtrsim \pi/4$, Eq. \eqref{deltawc} describes a metastable solution, with the stable solution obtained from it by the
replacement $\alpha\rightarrow \pi/2-\alpha$,  and change in sign of the r.h.s. of Eq.~\eqref{deltawc}. This is depicted in Fig. \ref{phasediag}. The nematic order dominates around $\alpha\sim 0$ with a similar
interaction dependence $\ln(W/\mathcal M)\sim 1/U\rho_0$, but with a more complicated full expression. These two orders possess the same ground state energy for $\sin^2(2\alpha)=2/3$, as predicted by the susceptibilities.
However, the nematic order parameter always coexists with a secondary, parasitic QAH order for the stable solution, satisfying $\Delta\sim U\rho_0 \mathcal M$ in the extreme weak coupling, $U\rho_0\ll 1$ limit.
 With increasing interaction, this
coexistence region shrinks and the region with pure QAH as the primary order parameter gains in territory.

A typical evolution of ordering with $\alpha$ is depicted in Fig.~\ref{phasediag} in the weak coupling limit. The QAH and nematic phases coexist only for the  stable solution, and exclude each other in the metastable
solutions of the first order transition.
The $\Delta=\mathcal M=0$ case always represents an unstable solution to the gap equations.
The full phase diagram, visualizing both stable and metastable regions, is plotted schematically in Fig.~\ref{phasediagall}, constructed from the numerical solution of the gap equations.
As predicted by the RG, there is a wide region for a QAH and nematic (accompanied by a subdominant QAH) phases. With increasing $U$, the region of the nematic
state shrinks, and eventually for large $U$ through a quantum phase transition gives way to the pure QAH state, in accordance with the considerations at strong
coupling.

\emph{Discussion.} The dice lattice with unequal hoppings and distinct sublattice potentials can be realized experimentally in a controlled way with
cold atoms loaded in an optical lattice\cite{raoux}. The interaction strength is tunable by e.g. a Feshbach resonance and by tuning the parameter
$\alpha$, so that our predictions can directly be tested.
In condensed matter, the dice lattice arises from a trilayer structure of the face-centred cubic lattice, grown in the [111] direction\cite{diracdora}, with 
SrTiO$_3$/SrIrO$_3$/SrTiO$_3$ trilayer heterostructures\cite{wangran} promising in this respect. Finally, the dice lattice can also be created by generalizing artificial graphene's honeycomb lattice\cite{manoharan}.

Cold atomic settings, unlike solid state ones, can host metastable states with a long lifetime due to the excellent control over various relaxation channels, offering the possibility to explore the full phase diagram.
Since both nematic and QAH states are gapped, a near-adiabatic tuning of $\alpha$  allows for passing through the stable to the metastable region.
Moreover, when two copies of our low-energy Hamiltonian, one for each valley, are realized by a given lattice model, a stable QAH state in one valley
and a metastable one in the other valley would always realize a metastable topological
insulating phase, which could be destroyed by slightly perturbing the system, giving way to charge density modulation.

\begin{acknowledgments}

Illuminating discussions with E. Szirmai on the RG are gratefully thanked.
BD is supported by the Hungarian Scientific Research Fund
Nos. K101244, K105149, K108676, CNK80991, ERC Grant Nr. ERC-259374-Sylo
and by the Bolyai Program of the HAS. IFH is supported by the NSERC of Canada.
\end{acknowledgments}

\bibliographystyle{apsrev}
\bibliography{refgraph}

\end{document}